\documentclass[twocolumn,showpacs,preprintnumbers,amsmath,amssymb,floatfix]{revtex4}

\usepackage{graphicx}
\usepackage{dcolumn}
\usepackage{bm}

\begin{document}

\preprint{APS/123-QED}

\title{Bose-Einstein Condensation in Competitive Processes}

\author{Hideaki Shimazaki}
\author{Ernst Niebur}%
\affiliation{%
Department of Neuroscience, Zanvyl Krieger Mind/Brain Institute, School of
Medicine, Johns Hopkins University, Baltimore, Maryland 21218
}%

\date{\today}

\begin{abstract}
We introduce an irreversible discrete multiplicative process that
undergoes Bose-Einstein condensation as a generic model of
competition. New players with different abilities successively join
the game and compete for limited resources. A player's future gain is
proportional to its ability and its current gain. The theory provides
three principles for this type of competition: competitive exclusion,
punctuated equilibria, and a critical condition for the distribution
of the players' abilities necessary for the dominance and the
evolution. We apply this theory to genetics, ecology and
economy. 
\end{abstract}

%
\maketitle 
Competition occurs when two or more players such as organisms,
individuals or companies strive for common but limited resources. It
plays a significant role in biological and social activities, and is
the basic idea underlying evolution. Recently, Bianconi \&
Barab\'{a}si demonstrated that their competitive complex network model
behaves analogously to a Bose gas \cite{Bianconi01}. A few nodes
recruited many connections and acted like hubs, while most others had
sparse links. This phenomenon was described in the mathematical
framework of Bose-Einstein condensation (BEC). Here, we analyze a
discrete multiplicative process that also exhibits BEC. This generic
competition model is defined by the following
conditions at each time step: 1. A player's gain is proportional to
its ability and to its gain at the previous time step. 2. A new player
joins the competition. 3. Players compete for a fixed total amount of
resources. The constant addition of players is the basis of the
model's open and irreversible characteristics and makes the process
suitable for the analysis of natural and social competition. We show
that the cumulative gain of each player is related to the Bose
distribution, except for that of two players. One is the pioneer. The
other is a high-ability player that dominates most resources under a
certain statistical condition for the distribution of the players'
ability. We investigated the implication of BEC in this mathematical game for
different competition models. Applied to the competition among clonal
strains of evolving asexual bacteria, BEC in an irreversible process
provides theoretical grounds for the principle of competitive
exclusion \cite{Gause64} and the hypothesis of punctuated equilibria
\cite{Elena96}. Furthermore, the existence of a transition point
elucidates the condition necessary for evolutionary development,
and explains the observation that a high mutation rate is not
necessarily advantageous for adaptive evolution
\cite{Sniegowski97,deVisser99}. Finally, we investigate possible
applications to ecological and economic activities.

Before we present our competition model, we introduce a novel, general
framework for how a function $\varphi(\epsilon)$ formally maps into a Bose
system. Define a function $\varphi(\epsilon)$ that satisfies the
following conditions for an arbitrary $C^2$ density function $g\left(
\epsilon \right) (\geq 0)$ defined on $\epsilon \in
[0,\epsilon_{\max}]$,
\begin{eqnarray}
 \int {d\epsilon \, g\left( \epsilon  \right)}  &=& 1, \label{33} \\
 \int {d\epsilon \, g\left( \epsilon  \right)\varphi \left( \epsilon  \right)}
&=& m + y_0, \label{34} \\
 \int {d\epsilon \, g\left( \epsilon  \right)e^{ - \beta \epsilon } \varphi
\left( \epsilon  \right)}  &=& M . \label{35}
\end{eqnarray}
The r.h.s. of these equations and $\beta$ in (\ref{35}) are all
positive constants. Then $\varphi \left( \epsilon  \right) $ is given by
\begin{equation}\label{36}
    \varphi \left( \epsilon  \right) = \frac{y_0}{{1 - e^{ - \beta
\epsilon  - \alpha } }},
\end{equation}
where $e^{-\alpha}   = {m \mathord{\left/ {\vphantom {m M}} \right. \kern-\nulldelimiterspace} M}$. For, dividing (\ref{34}) by $y_0$, then subtracting (\ref{33}) yields
\begin{equation}\label{24}
    \int {d\epsilon \, g\left( \epsilon  \right)\frac{{\varphi \left( \epsilon
\right) - y_0 }}{{y_0 }}}  = N,
\end{equation}
where $N = {m \mathord{\left/ {\vphantom {m {y_0 }}} \right. \kern-\nulldelimiterspace} {y_0 }}$. From (\ref{35})$\times {N \mathord{\left/ {\vphantom {N M}} \right. \kern-\nulldelimiterspace} M} - $(\ref{24}) and the fundamental lemma of the calculus of variation, we obtain (\ref{36}). Hence, the so-called occupation number in (\ref{24}),
\begin{equation}\label{37}
    n\left( \epsilon  \right) = \frac{{\varphi \left( \epsilon \right)
- y_0 }}{{y_0 }} = \frac{1}{{e^{\beta \epsilon  + \alpha } - 1}} ,
\end{equation}
becomes the Bose distribution, or, equivalently, the system $\varphi \left( \epsilon \right)$ maps into a Bose system. From (\ref{36}), such a system may be obtained from the sum of a geometric progression with ratio $e^{- \beta \epsilon - \alpha } < 1$. This motivates analysis of the following competition process.

Consider a simple multiplicative process
\begin{equation}\label{1}
    y_i(t+1) = a_i b(t)y_i(t),
\end{equation}
where $y_i(t)$ is the gain of the $i$th player at time $t$ and $a_i$ is its ability, a positive and time-independent random variable chosen from a distribution $\rho \left( a \right)$. From (\ref{1}), a player's gain is proportional to its ability and the gain of the previous time step. Furthermore, a new player joins the race at every time step with initial value $y_i(i) = y_0 $, which means we consider $t + 1 $ simultaneous equations at time $t $. The only exception is the first player (pioneer), whose initial value is $y_0 + m$. Finally, the population races for a fixed amount of resources, modelled by the normalization $b(t)$ in (\ref{1}) which is defined by $b(t) = {m \mathord{\left/ {\vphantom {m {M_t }}} \right. \kern-\nulldelimiterspace} {M_t }}$, where  $M_t$ is given by 
\begin{equation}\label{2}
M_t =  \sum\limits_{j = 0}^t {a_j y_j \left( t \right)}  .
\end{equation}
This assures that the total gain distributed among $t$ players at time
$t$ is $m $ for $t > 0$. As a new player joins the race with $y_0 $,
the total gain of the population is fixed to $m + y_0 $ every time
step. We now consider the time evolution of players except for the pioneer. The gain of the $i$th player at time $t$ is given by $y_i(t) =\Omega _i(t) y_0$, where
\begin{equation}\label{3}
    \Omega _i(t)  = a_i^{t-i} \left( {\prod\limits_{t' = i}^{t - 1}
{b\left( t' \right)} } \right) .
\end{equation}
Without loss of generality, we can set $a_i \leq 1$ (discussed later),
 and we let $a_i = e^{ - \beta \epsilon_i } $, thus introducing a
 formal parametrization of the ability distribution by the inverse
 temperature $\beta (=1/T) $. The random variable $\epsilon \in
 [0,\epsilon_{\max}]$ is chosen from $g\left( \epsilon \right)$,
a state density function which defines the system.
From (\ref{3}), we have
\begin{equation}\label{38}
 \ln \Omega _i = \left( {t - i} \right)\left\{ { - \beta \epsilon_i  -
\left\langle {\ln b} \right\rangle } \right\} ,
\end{equation}
using $\ln \prod\nolimits_{t' = i}^{t - 1} {b\left( t' \right)}  =
\sum\nolimits_{t' = i}^{t - 1} {\ln b\left( t' \right)}  \sim \left( t -i
\right) \left\langle {\ln b} \right\rangle $. By assuming stationarity of
$b(t)$, discussed below, we let its time average be
\begin{equation}\label{5}
    \left\langle {\ln b} \right\rangle  =  - \alpha .
\end{equation}
The normalization factor $\alpha$ keeps the total gain constant throughout the dynamical process. Its role is analogous to that of the chemical potential of a quantum gas introduced for the conservation of particle number. From (\ref{38}) and (\ref{5}), we obtain $y_i(t) = e^{( { - \beta \epsilon_i  - \alpha }) \left(t-i \right)} y_0 $. Given $e^{-\beta \epsilon_i - \alpha} < 1$, the cumulative gain $\varphi _t \left( \epsilon_i  \right) = \sum\nolimits_{t' = i}^t {y_i \left( {t'} \right)} $, converges to $\varphi \left( \epsilon_i \right)$ in (\ref{36}). Hence, the normalization of $\varphi _t \left( \epsilon_i  \right)$ by the initial value $y_0 $,
\begin{equation}\label{10}
n_t \left( \epsilon_i  \right) = \frac{{\varphi _t \left( \epsilon_i
\right) - y_0 }}{{y_0 }} ,
\end{equation}
takes the form of the Bose distribution in the thermodynamic limit $\left( {t \to \infty } \right)$. Indeed, in this limit, $M_t $ obeys
\begin{equation}\label{11}
 \mathop {\lim }\limits_{t \to \infty } M_t  = \int {d\epsilon \, g\left(
\epsilon  \right)\sum\limits_{t' = 0}^\infty  {e^{ - \beta \epsilon }
y\left( {t'} \right)} }  .
\end{equation}
Since the stationarity of $b\left( t \right)$ yields  $\mathop {\lim }\limits_{t \to \infty } \ln b\left( t \right) = \left\langle {\ln b} \right\rangle $, we have $\mathop {\lim }\limits_{t \to \infty } {m \mathord{\left/ \right. \kern-\nulldelimiterspace} {M_t }} = e^{ - \alpha } $ from (\ref{5}). By substituting this into (\ref{11}), we obtain an equality,
\begin{equation}\label{13}
\int {d\epsilon \, g\left( \epsilon  \right)\frac{1}{{e^{\beta \epsilon  +
\alpha }  - 1}}}  = N ,
\end{equation}
where $N = {m \mathord{\left/ {\vphantom {m {y_0 }}}
    \right. \kern-\nulldelimiterspace} {y_0 }}$. The dynamical system
(\ref{1}) thus formally maps into a Bose system in the thermodynamic limit. 

Comparing this system to a quantum gas, adding the gain $m$ at each
step corresponds to adding $m$ Bose particles. This gain is then
distributed among players with different abilities; these abilities
correspond to energy levels. Different from a quantum gas, a new
energy level is introduced at every step (with initial occupation of
$y_0$), the gains (particles) do not move to different energy levels
once they are allocated, and they are not quantized. Despite these
differences, the system preserves a basic feature of a
Bose gas: gains tend to accumulate where other gains are already
present. Thus, condensation analogous to BEC is expected at low
temperatures.

\begin{figure}[t]
\includegraphics[width=8.6cm]{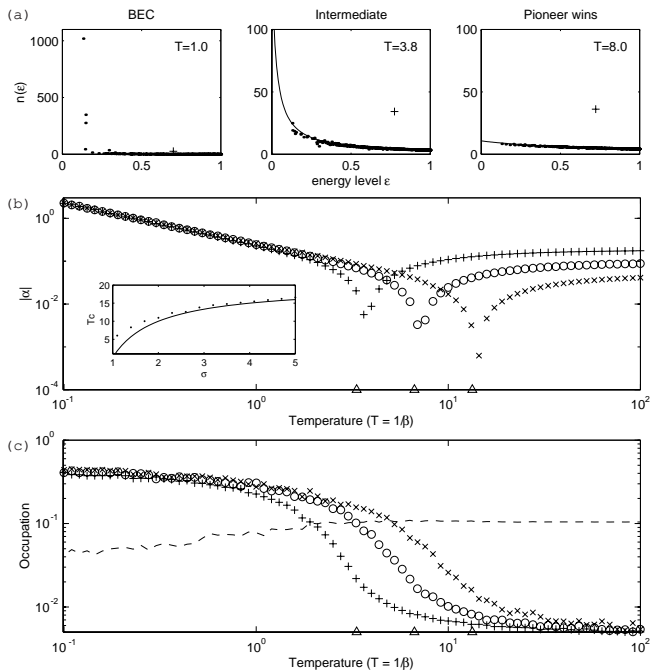}
\caption{\label{FIG1}  (a) Normalized cumulative gain (\ref{10}) for $m = 5$ and $T<T_c$ (left), $T \approx T_c$ (center) and $T>T_c$ (right). Solid line is the Bose distribution with $\alpha $ from (\ref{5}). The plus sign indicates a pioneer. Plots of the last ten entrants were excluded. (b) Numerical calculation of $|\alpha|$ with $\alpha$ from (\ref{5}), averaged over 500~trials. Symbols +, o, x indicate $m = 5$, $10$ and $20$.  Change in the sign of $\alpha$ indicates the transition. Triangles on the abscissa are the transition temperatures from (\ref{32}); $T_c \approx 3.33 $, $6.67 $ and $13.3 $. Shown in inset is the dependence of $T_c$ on exponent $\sigma $ for $m = 20$ (dots) and the analytical solution from (\ref{32}), solid line. (c) Cumulative occupation by the most capable player was calculated by $ \varphi_t \left( \epsilon _{\min } \right) \left/ \sum\nolimits_j \varphi_t \left( {\epsilon _j } \right) \right. $, where $\epsilon_{\min} = \min \left\{ \epsilon_j \right\} $ and $t = 200$. Symbols as in (c). Dashed line, occupation by pioneer for $m = 20$. In all simulations, $y_0 = 1 $, $\sigma = 3$ and $\epsilon _{\max } = 1$.}
\end{figure}

\begin{figure}[b]
\includegraphics[width=8.6cm]{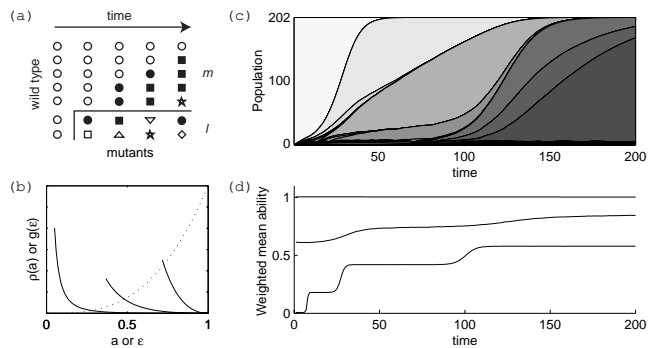}
\caption{\label{FIG2} (a) Schematic diagram of propagation of \textit{E. coli}. Symbols, o:~wild type (WT), others: mutants. (b) State density (\ref{39}) (dashed line) with $\sigma = 3$ and $\epsilon_{\max}=1$ used in simulation and corresponding ability distribution (solid line) $\rho \left( a \right) = g\left( \epsilon \right)\left| {{\textstyle{{d\epsilon } \over {da}}}} \right| $ for $T = 1/3, 1$ and 3 from left to right. Note $a \in [e^{ - \beta \epsilon _{\max } } ,1] $. (c) Time evolution at $T = 1$ ($m=200$, $l=2$, $y_0=1$).  As $T_c \approx 66.6$, the process is in BEC. Graded areas indicates population of each strains. WT abilities were set $\epsilon = 0.5$, which is slightly lower than $\left< \epsilon \right> \approx 0.76$ for $\sigma = 3$ since WT is likely to fit to environment better than chance. (d) Weighted mean ability shows punctuated equilibria at $T=0.1$ (bottom) and $T=1$ (center). At $T=100>T_c$ (top), stationarity (no evolution) is observed.}
\end{figure}

\begin{figure}[t]
\includegraphics[width=8.6cm]{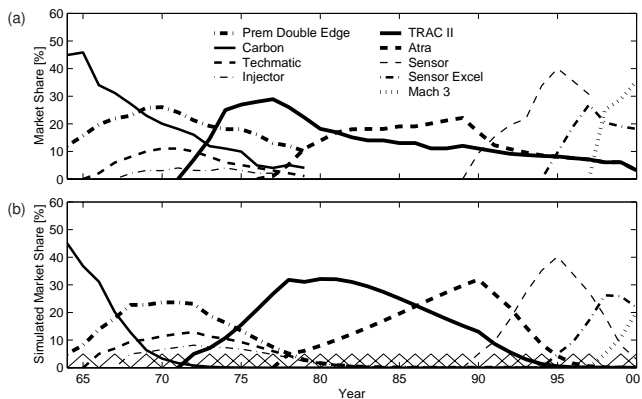}
\caption{\label{FIG3} (a) Market share of all nine different razors
  from The Gillette Company introduced over the last 3.5 decades
  \cite{Tellis01}.  (b) Simulated market share from (\ref{1}), using
  $y_0 = 1, m = 9$. Rather than assuming a specific state density, we
  assigned $a = 0.025, 0.04$, $0.043$ ... \textit{etc.} for successive
  products, with each newly introduced product having higher ability
  than the previous one. The specific values were selected to best fit
  the empirical data. In years without a product introduction by
  Gillette, the simulation introduced ``products'' with randomly drawn lower
  abilities ($a \ll 0.025$) that are not competitive and disappear
  (traces adjacent to abscissa). Share of the $j$th product at time
  $t$ was defined by ${y_j(t)} \left/ (m + y_0) \right. \times 45$ as
  45\% is the mean total share in (a). We considered the Carbon razor
  as the pioneer because it was the dominant product at the beginning of
  the period for which we have data. The introduction year of ``Prem
  Double Edge'' was not available, thus we assigned 1963 arbitrarily
  to fit the plot.  }
\end{figure}

To study this prediction, we simulated the multiplicative process
(\ref{1}), adopting  a standard  density function, 
\begin{equation}\label{39}
g\left( \epsilon  \right) = C_\sigma \epsilon ^{\sigma  - 1} ,
\end{equation}
where $C_\sigma = {\sigma \mathord{\left/
\right. \kern-\nulldelimiterspace} {\epsilon _{\max }^\sigma }}$
$\left( \sigma > 1\right)$.  We found two distinct phases for the
distribution of normalized cumulative gain (\ref{10}).  At high
temperature, $n_t \left( \epsilon_i \right)$ obeys the Bose
distribution (FIG.~\ref{FIG1}-a). Our assumption of stationarity of $b
\left( t \right)$ is confirmed numerically since $n_t \left(
\epsilon_i \right)$ agrees well with the Bose distribution with
$\alpha$ being provided on the assumption of stationarity. At low
temperatures, the coordinated distribution breaks down: one player
with low energy (but not necessarily the lowest energy because of the
normalization) dominates a large fraction of the cumulative
resources. Note that the stationarity of $b(t)$ no longer holds in
this domain. This Bose-Einstein condensate exists only below the
critical point $\alpha = 0$. For, if $\alpha $ is negative, $\varphi
\left( \epsilon_i \right)$ of the player with $\epsilon_i < - {\alpha
\mathord{\left/ \right. \kern-\nulldelimiterspace} \beta } $ does not
converge. Its normalized cumulative gain cannot be incorporated in the
integral in (\ref{13}); rather it has to be added as an extra
term. The critical temperature for this transition can be predicted in
the usual way. Change of variables $y = \beta \epsilon $ and use of
$\alpha = 0$ in (\ref{13}) yields
\begin{equation}\label{15}
N = \frac{\sigma }{{\left( {\beta_c \epsilon _{\max } }
\right)^\sigma  }}\int_0^{\beta_c \epsilon _{\max } } {\frac{{y^{\sigma  -
1} }}{{e^{y }  - 1}} \,dy} .
\end{equation}
By approximating the upper limit of integral in (\ref{15}) by infinity, the critical temperature $T_c({ = 1/\beta _c })$ is given by
\begin{equation}\label{17}
T_c^{\left( 1 \right)}  \sim  \epsilon _{\max } \left\{ {N^{ - 1} \Gamma
\left( {\sigma  + 1} \right)\zeta \left( \sigma  \right)} \right\}^{ -
{\textstyle{ \frac{1}{\sigma} }}} ,
\end{equation}
for $N \ll 1$. When $N \gg 1$, this approximation is not valid because of the high critical temperature. For $N \gg 1$, one uses $e^y  \simeq 1 + y$ in (\ref{15}) and obtains
\begin{equation}\label{32}
T_c^{\left( 2 \right)}  \sim  \epsilon _{\max } \left( {1 - \sigma ^{ - 1} }
\right)N.
\end{equation}
The transition can be easily seen in the occupation by the most
capable player, defined by ${\varphi_t \left( {\epsilon _{\min } }
\right)} \mathord{\left/ {\sum\nolimits_j {\varphi_t \left( {\epsilon
_j } \right)} } \right.} $, where $\epsilon_{\min} = \min \left\{
\epsilon_j \right\} $. As the temperature decreases below $T_c $, its
occupation dramatically increases, supporting the prediction
(FIG.~\ref{FIG1}-c). The time evolution of each player's gain also
shows characteristic differences above and below the predicted
$T_c$. At $T > T_c$, the gain of each player monotonically
decreases. Below $T_c$, time evolution is not monotonic
(FIG.~\ref{FIG3}-b and area in FIG.~\ref{FIG2}-c). This is
particulary interesting because this BEC occurs in an irreversible
process. In contrast to BEC in a quantum gas in equilibrium, the
condensation state dose not imply the dominance of resources by a
single player at the lowest energy throughout the whole competition
process. Instead, there are replacements of dominant player upon the
entrance of a player with lower energy level.


We verified the existence of a BEC analogue in a discrete
multiplicative process, as was shown in a continuous model
\cite{Bianconi01}. To see its applicability to real-world competition,
we mention that (\ref{1}) is a natural extension of Malthusian
population increase: population, when unchecked, increases in a
geometrical ratio, that is $y_i \left( {t + 1} \right) = \tilde a_i
y\left( t \right)$, where $\tilde a_i \in \left[ {\tilde a_{\min }
    ,\tilde a_{\max } } \right]$ and $\tilde a_{\min } > 1$.  We
introduced normalization to model limited resources, obtaining
(\ref{1}) with $a_i$ replaced by $\tilde a_i$. Note that we can
substitute $\tilde a_i = \tilde a_{\max} a_i$ (i.e. $a_i \leq 1$)
without loss of generality  because $\tilde a_{\max}$ is common in the
nominator and the denominator in  (\ref{1}). This enables us to apply
(\ref{1}) to population dynamics and even economic activities.

\textit{Application to Genetics:} We consider competition of clonal
strains of asexual \textit{Escherichia coli} serially propagated on
glucose-limited medium. For this aim, we slightly generalize model
(\ref{1}) such that we allow $l \left( \geq 1 \right)$ new players
to join the competition with initial value $y_0$
at each step. From a similar argument, we obtain (\ref{13}), where $N
= m \left/ y_0 l \right.$.  Assigning $y_0 = 1$, we have $l$ mutants
(with usually different abilities) at each time step
\cite{Footnote1}. We chose this procedure because it is unlikely to
have $y_0 > 1$ mutants of the same ability. The original $m + y_0 l$
wild types propagate $m$ wild types and produce $y_0 l$ mutants of
novel abilities. Then the mixed population of $m + y_0 l$ is
normalized to $m$, and generates $y_0 l$ mutants (FIG.~\ref{FIG2}-a,
c). Eventually all wild types are swept away and replaced by
mutants. This is much faster in BEC than at high temperatures (see
reduction of pioneer occupation in BEC in FIG.~\ref{FIG1}-c). Also at
low temperatures inferior strains are quickly expelled by a superior
strain, which is analogous to the competitive exclusion principle
\cite{Gause64,Gerrish98}. Subsequent transient replacement of the
dominant mutants supports the hypothesis of punctuated equilibria
\cite{Elena96}.  Weighted mean ability ($M_t$ in (\ref{2}) if $l =1$)
shows step-like evolution at low temperatures (FIG.~\ref{FIG2}-d), but
this effect diminishes with increasing temperature. Now, new mutants
all have similar abilities, competition is crowded and severe, and the
sudden upsurges of the weighted mean ability of the population are
smeared out. Finally, at sufficiently high temperatures, no mutant
becomes an absolute winner; evolution ceases when the population lacks
prominent players.  Note that the stationarity of weighted mean
ability confirms the stationarity of $b(t)$, which in turn allows the
argument below (\ref{5}). 

A striking statement of the theory is the prediction of a singular
point on the emergence of evolution. Most mutations are likely to be
deleterious \cite{Gerrish98}, which was simulated by the use of an
increasing state density (\ref{39}) so that inferior mutants are
likely to occur (FIG.~\ref{FIG2}-c). In this case, the critical
temperature is predicted by (\ref{32}). Through its dependence on $N$,
$T_c$ is found to be dependent on the mutation rate ${{y_0 l}
\mathord{\left/ \right.  \kern-\nulldelimiterspace} {\left( {m + y_0
l} \right) = \left( {N + 1} \right)^{ - 1} }} $. Therefore, at a given
temperature, decreasing mutation rate generates evolutionary
development. A sufficiently low mutation rate not only prevents
deleterious offspring but is necessary for adaptive evolution
itself. 

Although introducing the notion of temperature and the state density
of a quantum gas
was instructive, our model  does not require these
constructs. Our conclusions hold as long as a sharply decreasing
ability distribution $\rho_\lambda(a)$ exists on $a \in [a_{\min},1]$ where $a_{\min}>0$, with $\lambda$ a parameter,  such that the integral (\ref{13}),
\begin{equation}\label{41}
I\left( {\lambda ,\alpha } \right) = \int_{a_{\min}}^1 {da \, \rho _\lambda  \left( a
\right)\frac{1}{{a^{ - 1} e^\alpha - 1}}}  = N ,
\end{equation}
is finite. Since $I \left( {\lambda ,\alpha } \right) \leq I\left(
{\lambda ,0} \right) \equiv N_c \left( \lambda \right) $, we have a
finite $N_c$ and we can determine the corresponding critical mutation
rate for a given $\lambda$. Since $N_c$ is finite, the process is in
condensation for $N > N_c $, while evolution ceases for $N < N_c$. In
those cases in which the critical mutation rate is in the
physiological range, this argument explains ``clonal interference''
\cite{Gerrish98} which is an empirically observed constraint on the
role of mutator alleles that rises to high frequency in association
with beneficial mutations \cite{Sniegowski97,deVisser99}. The critical
point, however, only determines the occurrence of adaptive evolution
at $t \rightarrow \infty$, and a low mutation rate slows the rising of
dominant mutants. This implies the existence of an optimal mutation
rate for a transiently changing environment. 

\textit{Application to Ecology and Economy:} In ecology, the
Lotka-Volterra equation predicts that coexistence
between strongly competing species is impossible, a result confirmed
in laboratory experiments on \textit{Paramecium} \cite{Gause64} 
though applicability in complex ecosystems is uncertain \cite{Colinvaux86}. 
These results established the \textit{principle of competitive
exclusion}; or ``one species, one niche.'' Our process models interspecific competition with completely
overlapping niches, and BEC is a natural extension of the competitive
exclusion principle in an irreversible process. This principle may
also apply to economic activities if a product or a company satisfies
the ecological conditions of the exclusion principle
\cite{Hardin60}. For instance, products sold for a specialized market
may compete within their niche in the marketplace. Figure \ref{FIG3}-a
shows the market share of different razors successively introduced by
The Gillette Company, illustrating a much-studied case of new products
eroding the market share of their own predecessors
\cite{Tellis01}. Aspects of the dynamics of product replacement may
thus be captured by our  competitive process (Fig.~\ref{FIG3}-b).

We thank HG Schuster for very helpful suggestions and the NSF for support through a CAREER award to EN.
\bibliography{bec}
\end{document}